\documentclass[namedreferences]{SolarPhysics}
\usepackage[optionalrh]{spr-sola-addons} % For Solar Physics 
\usepackage{graphicx}        % For eps figures, newer & more powerfull
\usepackage{color}           % For color text: \color command
\usepackage{url}             % For breaking URLs easily trough lines
\newcommand{\aap}{    {\it Astron. Astrophys.}}

\newcommand{\apj}{    {\it Astrophys. J.}}
\newcommand{\apjl}{   {\it Astrophys. J. Lett.}}

\newcommand{\mnras}{  {\it Mon. Not. Roy. Astron. Soc.}}

\newcommand{\solphys}{{\it Solar Phys.}}

\begin{document} 

\begin{article} 

\begin{opening} 

\title{Multiscale magnetic underdense regions on the solar surface: Granular and Mesogranular scales}
\author{F.~\surname{Berrilli}$^{1}$, S.~\surname{Scardigli}$^{1}$, S.~\surname{Giordano}$^{1,2}$
      } 
\runningauthor{Berrilli, Scardigli \& Giordano} 
\runningtitle{Multiscale magnetic underdense regions} 
\institute{$^{1}$ Department of Physics, University of Rome Tor Vergata, I-00133, Rome, Italy 
                    email: \url{berrilli@roma2.infn.it} \\  
             $^{2}$ ALTRAN SpA, Rome Area, Italy} 

\begin{abstract} 
The Sun is a non-equilibrium dissipative system subjected to an energy flow which originates in its core. Convective overshooting motions create temperature and velocity structures which show a temporal and spatial evolution. As a result, photospheric structures are generally considered to be the direct manifestation of convective plasma motions. The plasma flows on the photosphere govern the motion of single magnetic elements. These elements are arranged in typical patterns which are observed as a variety of multiscale magnetic patterns.\\
High resolution magnetograms of quiet solar surface revealed the presence of magnetic underdense regions in the solar photosphere, commonly called voids, which may be considered a signature of the underlying convective structure. The analysis of such patterns paves the way for the investigation of all turbulent convective scales from granular to global.\\ In order to address the question of magnetic structures driven by turbulent convection at granular and mesogranular scales we used a “voids” detection method.\\ The computed voids distribution shows an exponential behavior at scales between 2 and 10 Mm and the absence of features at 5-10~Mm mesogranular scales. The absence of preferred scales of organization in the 2-10~Mm range supports the multiscale nature of flows on the solar surface and the absence of a mesogranular convective scale.
\end{abstract} 
\end{opening} 
%------------------------------------------------- 

\section{Introduction}
    \label{S-Introduction}  

The Sun is a non-equilibrium dissipative system subjected to an energy flow which originates in its core. A turbulent convective envelope is generated by this energy flow in response to the significant opacity of the outer solar region. The photosphere shows a state of reduced spatial symmetry due to the convective overshooting motions on the surface. The result of this reduced symmetry is the formation of temperature and velocity structures which show a temporal and spatial multiscale evolution. As a result, photospheric structures are generally considered to be the direct manifestation of convective motions.\\ 
Global plasma flows (i.e., differential rotation, meridional circulation and torsional oscillation) and the solar magnetic field, interacting with these turbulent convective motions, increase the complexity of emerging structures and of the star as a whole. The dynamics of the solar surface and its interaction with the magnetic field ultimately control the  structure of the outer solar atmosphere and the heliosphere beyond.\\ 
The plasma flows on the photosphere govern the motion of single magnetic elements. These elements are arranged in typical patterns which are observed as a variety of multiscale magnetic features, e.g., filamentary clusters or clumps of magnetic elements, magnetic network, clusters of facular points, active regions.
The analysis of these multiscale magnetic patterns on the solar surface paves the way for the investigation of all turbulent convective scales from granular to global. \opencite{Chaouche} report, indeed, that {\it the magnetic field can provide a direct avenue to explore convective patterns since the magnetic flux can be measured directly using Stokes polarimetry techniques}. \\ 
Commonly, four different convective scales are identified on the solar surface: the granulation, which shows a typical horizontal scale of about 1 Mm and a short (a few minutes) lifetime, the mesogranulation, which shows horizontal length scales ranging from 5 Mm to 10 Mm and a lifetime of some hours, the supergranulation, which shows typical horizontal length scales from 20 Mm to 50 Mm and a lifetime of about one day, and giant cells which show horizontal length scales of the order of 100Mm or larger.\\
The wide range of values which are reported in the observed horizontal length scales of mesogranulation and supergranulation are most likely due to different methods of data analysis and choices of spatial and time filters.
Even more important, however, is that this division, as suggested in \opencite{Nordlund2009}, {\it is largely of historical origin, and current evidence indicates that there is a continuous spectrum of motions, on all scales from global to sub-granular}.\\

In the last years, high resolution magnetograms of quiet solar surface (e.g., \opencite{berger}, \opencite{almeida}, \opencite{lites08}) revealed the presence of multiscale magnetic underdense regions in the solar photosphere, commonly called voids, which may be considered a signature of the underlying convective structure. 

From magneto-convection simulations it results that at the granular scales the magnetic field elements are concentrated at the boundaries of granular cells (i.e., intergranular lanes). Several observations, i.e., \opencite{cerdena03a}, \opencite{Khomenko03}, \opencite{manso}, confirmed such numerical results. The presence of magneto-convective concentration at mesogranular scales has been predicted by numerical simulations (\opencite{Cattaneo01}) and observed with 2D spectro-polarimetry (\opencite{cerdena03b}, \opencite{Chaouche}), IR spectropolarimetry (\opencite{truillo03}) and from magnetograms (\opencite{almeida}).\\ 
In more detail, \opencite{cerdena03b} assessed the existence of a web-like pattern with the spatial scale of mesogranulation by using a time sequence of Inter-Network magnetograms observed at the solar disk center. The authors reported that the observed persistent pattern resembled a network with a spatial scale between 5 and 10 arcsec, which they identified as mesogranulation. 

More recently, \opencite{Chaouche} confirmed the preferential location of magnetic elements in mesogranular cells using high spatial and temporal resolution IMaX-SUNRISE dataset. The mesogranular cells were identified using Lagrange tracers driven by horizontal velocity fields which were computed via Local Correlation Tracking technique (\opencite{November88}). In addition, the statistical multiscale analysis of the photospheric field geometry outlined the role of solar surface convective motions in the clustering and intermittent organization of magnetic flux elements in the quiet Sun (\opencite{lawrence93},\opencite{Uritsky}).
Moreover, \opencite{deWijnM} reported that it seems highly likely that the cells described by \opencite{deWijn05} and the “voids” found by \opencite{lites08} correspond to mesogranules.\\

In order to address the question of which organization scales appear in magnetic patterns formed by turbulent convective motions and how these scales could be revealed in high resolution magnetograms, we used a “void” detection method. Void detection methods are largely used in cosmology to study the voids from galaxy and clusters (\opencite{aikio98}). These numerical methods define void structures of different particle distributions. In these methods galaxies are considered single {\it particles}. High-density particle regions surrounding low-density regions actually form the edges of these empty volumes, called voids.
In the case of magnetic patterns on the solar surface we observe areas of low or absent magnetic activity surrounded by intense magnetic structures. Like the voids in the distribution of galaxies we define this underdense magnetic field regions in solar magnetograms as voids.\\

In this work we employ a suitable automated void detection procedure to single out these underdense regions and to investigate their geometrical properties. We concentrate our analysis on the remarkable scales of granulation, the only length scale for which the convective signature is directly visible on the solar surface, and mesogranulation, whose physical origin is still under debate. See \opencite{Nordlund2009} for a recent review on solar surface convection. In more detail, we are concerned with the study of distribution of voids in a quiet Sun high resolution magnetogram observed at disk center.\\
%Our conclusions are given in Section~\ref{S-Conclusion}. 

%-------------------------------------------------------------
\begin{figure}[h!]
\includegraphics[width=1\textwidth,clip=]{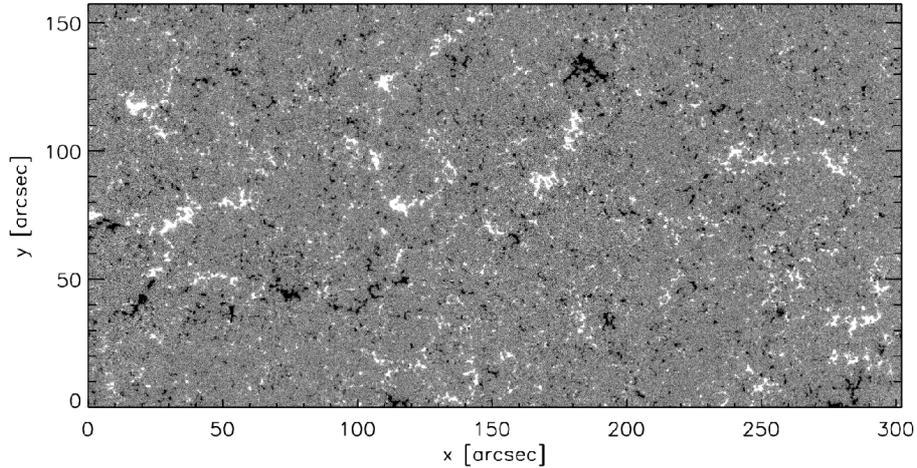}
\caption{Hinode SOT/SP line-of-sight magnetogram of a $302 \times 162$ $arcsec^{2}$ portion of the solar photosphere observed at disk center. }
\label{magnetogram}
\end{figure}
%-------------------------------------------------------------
%Our analysis is based on a single high resolution magnetogram (saturated at 200G) obtained via the COG method. 
\section{Dataset}
\label{S-dataset} 
We analyzed a high resolution magnetogram of a $302 \times 162$ $arcsec^{2}$ portion of the solar photosphere observed at disk center on 2007 March 10 between 11:37 and 14:34 UT (Fig.\ref{magnetogram})
The spectropolarimetric measurements were taken by the Solar Optical Telescope SOT/SP instrument aboard HINODE (\opencite{tSuneta}). The magnetogram has been derived from the FeI 630.15~nm  Stokes I and V signals by adopting the center-of-gravity method (\opencite{reessemel}).\\
We adopted the plate scale reported in the FITS header information of the Hinode SP, i.e., 0.1476$arcsec$/px and 0.1585$arcsec$/px along x and y, respectively. The (diffraction-limited) spatial resolution is about 0.32 $arcsec$. See \opencite{lites08} for a detailed description of the dataset.

\section{Void detection algorithm and results}
\label{S-algorithm}  
In this work we adopted a 2-D version of the algorithm developed by \opencite{aikio98} to detect voids in three dimensional red-shift galaxy surveys. 
Below, we formally define the void by analogy with \opencite{Einasto89} and \opencite{aikio98}.\\

Let us consider a distribution of particles in a square region L$^2$ $\in$ $\Re^2$ with side length $L$. We define a scalar field, called \textit{distance field} (DF), $D:L^2\longrightarrow R$ as the distance of a given point $x$ in L$^2$ to the nearest particle:
%***************************
\begin{equation} 
D(x)=\min _n \left\{\mid x-X_n\mid\right\}
\label{eq6-1}
\end{equation} 
%***************************
where $X_n$, n=1,. . ., N are the positions of the particles.\\
The local maxima of the distance field, which are the points with the longest distance to the nearest particle, are the ''centers'' of empty regions, i.e., voids. 
To compute the DF we superimpose on the region $L^2$ a two-dimensional grid of $k^2$ elementary cells of equal area, where $k=L/s$. The parameter $s$, called \textit{resolution parameter}, sets the spatial resolution of the void analysis. For each cell center we compute the distance to the nearest particle, thus obtaining the discrete DF $D(x)$.\\ 
Cells corresponding to the $i_{th}$ local maximum of $D(x)$ are identified with $M_i$. Subvoid $v_M$ $\in$ $L^2$ is a region around $M_i$ containing all the points $x$ continuously connected to the maximum $M_i$ along a monotonically increasing $\nabla D(x)$ path. More than one maximum of $D(x)$ could fall inside a void; such local maxima belong to subvoids of the same void. 
Although a subvoid clearly belongs to a determined void, the subvoid does not necessarily cover the whole void area. 
%Let the distance between maxima $M_1$ and $M_2$ be $d_{1,2}=\mid M_2-M_1\mid$. 
In general we say that subvoids $v_{M_i}$ and $v_{M_j}$ belong to the same void $V$ if: 
\begin{enumerate}
\item the maxima $M_i$ and $M_j$ are nearer to each other than the nearest particle 
%****************************
\begin{equation} 
d_{i,j}\leq \min \left\{D(M_i),D(M_j)\right\}
\label{eq6-2}
\end{equation} 
%****************************
where $d_{i,j}$ is the distance between $M_i$ and $M_j$ maxima, and $D(M_k)$ is the value of the $D(x)$ at point $M_k$ and ; 
\item if there exists a chain of subvoids $\left\{{v_{M_k}}\right\}_{k=1,n}$ such that:\\ 
$d_{i,1}\leq  \min\left\{D(M_i),D(M_1)\right\}$, $d_{k,k+1}\leq \min\left\{D(M_k),D(M_{k+1})\right\}$\\ 
for all k=1,. . . , n-1 and  \\
$d_{n,j}\leq \min\left\{D(M_n),D(M_{j})\right\}$ (i.e. there exists a chain of subvoids obeying the condition in Eq.\ref{eq6-2}).
\end{enumerate}
%-------------------------------------------------------------
\begin{figure}[h!]
\includegraphics[width=1\textwidth,clip=]{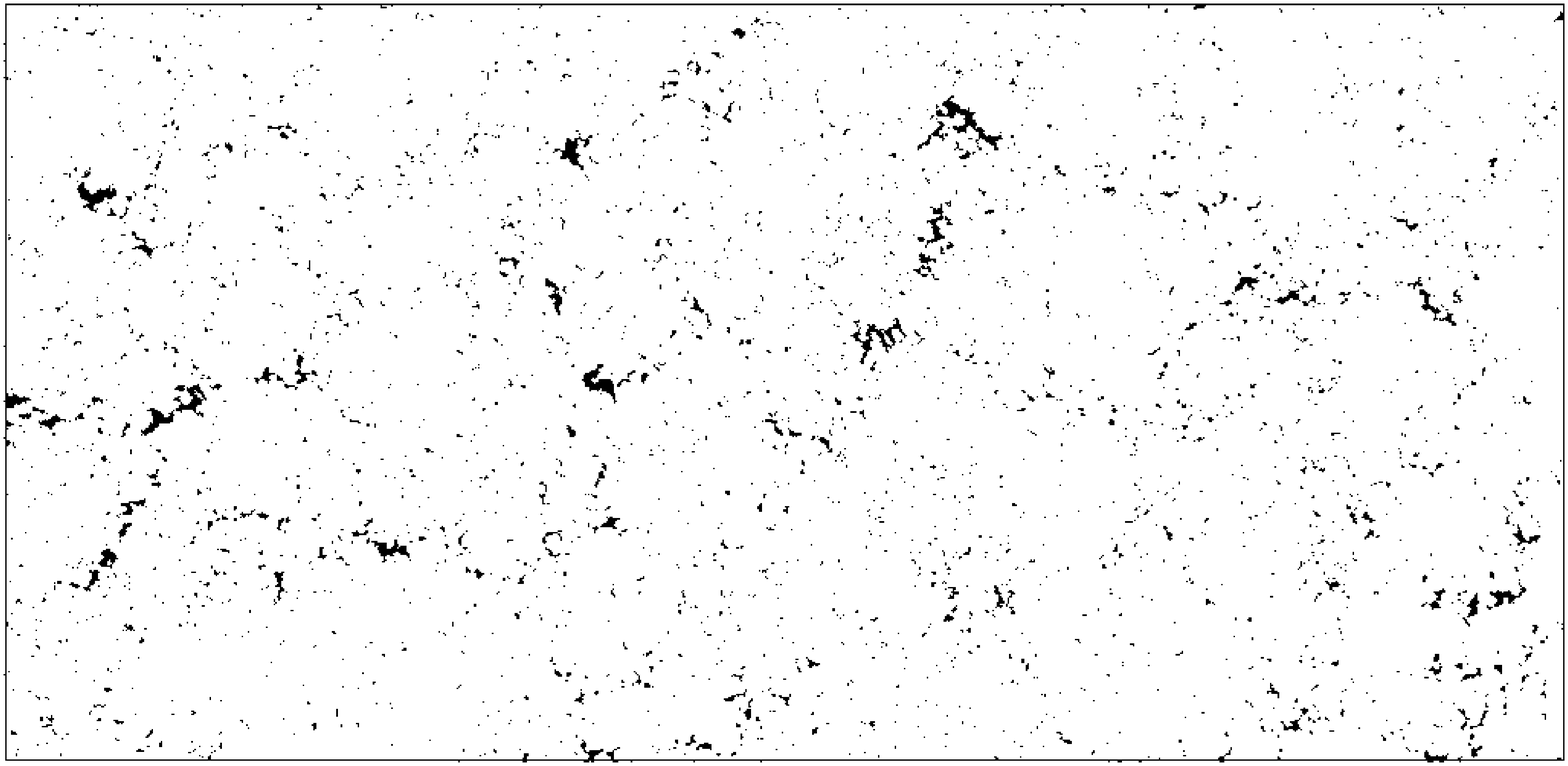}
\caption{Hinode SOT/SP line-of-sight two-levels magnetogram. The two-levels version of the magnetogram is computed taking into account only those pixels with magnetic absolute signal above the typical equipartition field strength of 200~G.}
\label{binary}
\end{figure}
%-------------------------------------------------------------
The output of the algorithm contains the number $N_c$ of elementary cells in each void and the main radius R$_V$ of the voids. The number of elementary cells defines the area of the void, in fact the area A is equal to $N_c$ multiplied the constant area $s^2$ of elementary cell, i.e., a pixel in our computational volume. 
The void length scale is computed as the equivalent diameter of a circular region with the same area, i.e., $2\times \sqrt{A/\pi}$, where A is the area of the detected void.\\
The main radius R$_V$ is the maximum of the $D(x)$ inside the void, i.e., it is the maximum of the $D(x)$ maxima D(M$_k$) inside the void:
\begin{equation} 
R_V=\max _{M_k\in V}\left\{D(M_k)\right\}
\label{eq6-3}
\end{equation} 

\begin{figure}[h!]
\includegraphics[width=1\textwidth,clip=]{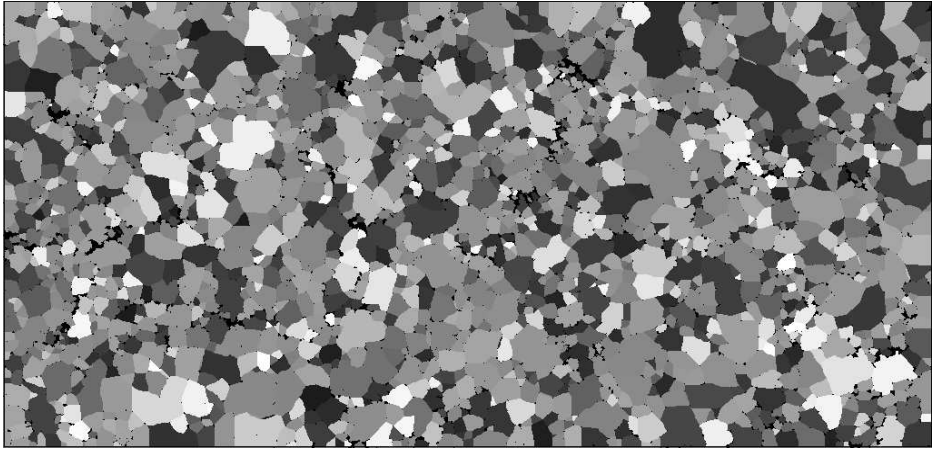}
\caption{Hinode SOT/SP line-of-sight segmented magnetogram. The image shows 1951 identified voids.}
\label{voids}
\end{figure}

\begin{figure}[h!]
\centering\includegraphics[width=0.65\textwidth,clip=]{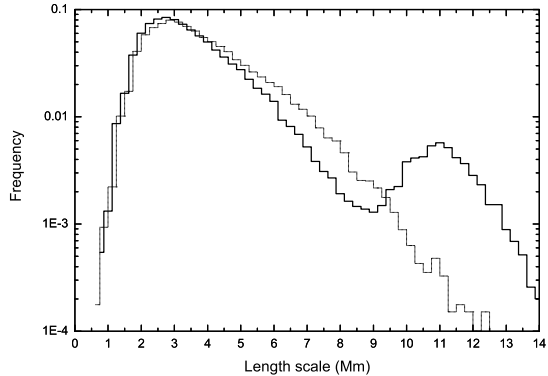}
\caption{Average void distributions for pseudorandom patterns of structures (dot) and pseudorandom patterns of structures with 100 round empty regions which generate voids of mesogranular scale (heavy continuum). In the second case, the final distribution shows two populations. The first population is identified by the 3~Mm maximum related to the correlation distance among structures. The second population is identified by the second maximum related to the mesogranular scale voids.}
\label{distributionS}
\end{figure}

To start the algorithm we used a global threshold to get the two-levels version of the HINODE magnetogram. Since the mechanism of flux expulsion, for photospheric convection, can concentrate magnetic flux up to the equipartition field strength, which is about 200 G at the solar surface (\opencite{thomas}, \opencite{solankietal}), we decided to set the threshold value to 200 Gauss (absolute value). Therefore, in order to compute the two-levels version of the magnetogram we took into account only those pixels with magnetic signal above the typical equipartition field strength of 200~G. 
In Fig.\ref{binary} we show the two-levels magnetogram.
After detecting the voids present in the magnetogram (Fig.\ref{voids}) we use an automated procedure to calculate area and position of each cell. The algorithm extracted 1951 voids. In Fig.\ref{distributionS} we show the distribution of void length scale.\\
%-------------------------------------------------------------
%-------------------------------------------------------------
%-------------------------------------------------------------%-------------------------------------------------------------

%-------------------------------------------------------------
We applied the algorithm to several test images in order to validate its reliability and to study void distributions under controlled conditions. 
Particularly, we applied the void searching procedure on two ensembles of 20 simulated patterns of pseudorandom structures. The size of the computational volume and the number of structures were set equal to the size of the Hinode magnetogram and to the number of observed magnetic features, respectively. The samples of the first ensemble did not include privileged size voids, while the samples forming the second ensemble included 100 pseudorandom round empty regions which generate voids of mesogranular scale.\\ 
Final void distributions, for both ensembles, were computed by considering all the voids of 20 samples (Fig.\ref{distributionS}).\\ 
The final distribution shows two populations only for pseudorandom patterns of structures with non-overlapping mesogranular voids. The first population is identified by the 3~Mm maximum related to the correlation distance among structures. The second population is identified by the second maximum and it is due to the presence of the included mesogranular voids.\\

\section{Discussion and conclusions}
\label{S-conclusions} 

\begin{figure}[h!]
\centering\includegraphics[width=0.65\textwidth,clip=]{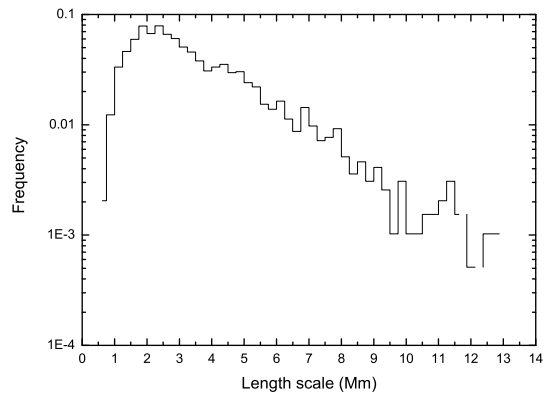}
\caption{Void frequency distribution of 1951 identified voids.}
\label{distributionHINODE}
\end{figure}

As reported in paragraph \ref{S-algorithm}, we applied the algorithm to a large SOT-HINODE magnetogram and the result is that 1951 magnetic underdense regions, i.e., voids, were identified.\\
The calculated  distribution of these 1951 voids (Fig.\ref{distributionHINODE}) shows an exponential decrease between 2 and 10 Mm with a decay constant equal to $2.2\pm0.2$ Mm and a coefficient of determination $R^2=0.98$. The magnitude of $R^2$ indicates a very high degree of correlation. The absence of features at scales of 5-10~Mm indicates the lack of an intrinsic mesogranular scale.\\
The effect of different threshold values is investigated from a comparison of corresponding void distributions. Our findings show that the decay constant decreases when the threshold increases from 120~Gauss to 260~Gauss. The distributions remain exponential with the same high degree of correlation, even though the mesogranular signature is not observed.\\
Thus, we support the results reported by \opencite{Chaouche} that mesogranulation is not among the primary energy-injection scales of solar convection. In addition, these authors report that the Probability Density Function of the distance between magnetic footpoints shows a constancy of the slope at scales between 1 and 10 Mm and that the characteristic decay distance is approximately 1.7 Mm. Moreover, they reported that only after about 20 minutes of integration a sharp mesogranular network appeared.\\ 
Similarly, \opencite{berrilli} found that mesogranular photospheric features appear on the solar surface with a characteristic growth time of about 10~minutes. They used the hexagonal normalized information entropy measure, $H'_{hex}$, to observe the mesogranular scale emergence and to compute the associated time of formation. They found that the $H'_{hex}$ signal shows a maximum on the mesogranular scale of 8~Mm after 15-20 minutes of integration.\\
%-------------------------------------------------------------

The absence of specific scales of organization in the 2-10~Mm range supports the multiscale nature of flows on the solar surface, but does not rule out the possibility of convection structures with these dimensions. For example, large scale flows can be obtained as the result of the local merge of downward plumes in advective-interaction models (\opencite{rast}).\\ 
Exponential functions describe other physical quantities in turbulent convective systems, e.g., the probability density of normalized temperature fluctuations shows an interesting feature at the transition from soft to hard turbulent convection (\opencite{yakhot}). As stated by authors, in the soft turbulent convective regime, when the Rayleigh number $Ra < 10^{7}$, the probability distribution measured at the center of the cell is Gaussian, while in the hard-turbulence regime ($Ra>10^{8}$) the probability distribution is very close to exponential. 
The value of the Rayleigh number characterizing intensity of the turbulent convection in photosphere is $Ra\simeq10^{11}$ (\opencite{Bray}) which is consistent with the hard-turbulence regime.\\

The hypothesis that there is a continuous spectrum of flows on the solar surface, from global to sub-granular (\opencite{Nordlund2009}), still needs to be thoroughly \begin{tiny}
•
\end{tiny}proven. With this in mind, we are developing a novel fast void detection algorithm to analyze a large dataset of MDI magnetograms acquired during the “exceptional” solar minimum at the end of solar cycle 23. The analysis of this dataset will allow us to investigate voids in the range of 5-100~Mm in order to validate the above hypothesis.

%% Acknowledgements
%
 \begin{acks}
We wish to thank Bartolomeo Viticchi\`e for his contribution in the analysis of the HINODE data and Michael Senno and Roberto Piazzesi for their contributions in reviewing the manuscript. This project is supported by the University of Rome Tor Vergata Astronomy Ph.D. Program. Hinode is a Japanese mission developed and launched by ISAS/JAXA, with NAOJ as domestic partner and NASA and STFC (UK) as international partners. It is operated by these agencies in cooperation with ESA and NSC (Norway).
 \end{acks}

%%%%%%%%%%%%%%%%%%%
%%% BIBLIOGRAPHY
%%%%%%%%%%%%%%%%%%%%%%%%%%%%%%%%%%%%%%%%%%%%%%%%%%%%%%%%%%%
\mbox{}~\\

\end{article} 

\begin{thebibliography}{}

\bibitem[\protect\citeauthoryear{{Aikio \& M\"ah\"onen}}{1998}]{aikio98}
Aikio J. and M\"ah\"onen P., 
1998, {\it{\apj}}, \textbf{497}, 534.

\bibitem[\protect\citeauthoryear{{Berger et al.}}{1998}]{berger}
Berger T. E., L\"ofdah M. G., Shin R. S., \& Title A. M. 
1998, {\it{\apj}}, \textbf{495}, 973.

\bibitem[\protect\citeauthoryear{{Berrilli et al.}}{2005}]{berrilli}
Berrilli F., Del Moro D., Russo S., Consolini G., Straus Th.,
2005, {\it{\apj}}, \textbf{632}, 677.

\bibitem[\protect\citeauthoryear{{Bray et al.}}{1984}]{Bray}
Bray R. J., Loughead R. E. and Durrant C. J.
1984, {\it The Solar Granulation} 2nd edition Cambridge University Press

%\bibitem[\protect\citeauthoryear{{Bershadskii}}{2009}]{Bershadskii}
%Bershadskii A.,
%2009, {\it Europhysics Letters}, \textbf{85}, 49002.

\bibitem[\protect\citeauthoryear{{Cattaneo et al.}}{2001}]{Cattaneo01} 
Cattaneo F., Lenz D., \& Weiss N., 
2001, {\it{\apj}}, \textbf{563}, L91.

\bibitem[\protect\citeauthoryear{{de Wijn \& M\"uller}}{2009}]{deWijnM} 
De Wijn A. G. \& M\"uller D.
2009, {\it ASP Conference Series}, \textbf{415}, 211.

\bibitem[\protect\citeauthoryear{{De Wijn et al.}}{2005}]{deWijn05}
De Wijn A. G., Rutten R. J., Haverkamp E. M. W. P., and Suetterlin P.
2005, {\it{\aap}}, \textbf{441}, 1183.

\bibitem[\protect\citeauthoryear{{Dom\'inguez Cerde\~na et al.}}{2003a}]{cerdena03a} 
Dom\'inguez Cerde\~na I., Kneer F., \& S\'anchez Almeida J., 
2003a, {\it{\apj}}, \textbf{582}, L55.

\bibitem[\protect\citeauthoryear{{Dom\'inguez Cerde\~na et al}}{2003b}]{cerdena03b}
Dom\'inguez Cerde\~na, I., S\'anchez Almeida, J., \& Kneer, F., 
2003b, {\it{\aap}}, \textbf{407}, 741.

\bibitem[\protect\citeauthoryear{{Einasto, Einasto \& Gramann}}{1989}]{Einasto89}
Einasto J., Einasto M., Gramann M., 
1989, {\it{\mnras}}, \textbf{238}, 155.

\bibitem[\protect\citeauthoryear{{Khomenko et al.}}{2003}]{Khomenko03} 
Khomenko, E. V., Collados, M., Solanki, S. K., Lagg, A., \& Trujillo  Bueno, J., 
2003, {\it{\aap}}, \textbf{408}, 1115.

\bibitem[\protect\citeauthoryear{{Lawrence, Ruzmaikin \& Cadavid}}{1993}]{lawrence93} 
Lawrence J. K., Ruzmaikin A. A., Cadavid A. C.
1993, {\it{\apj}}, \textbf{417}, p.805.

\bibitem[\protect\citeauthoryear{{Lites et al.}}{2008}]{lites08}
Lites B. W., Kubo M., Socas-Navarro H., Berger T., Frank Z., Shine R., Tarbell T., Title A., Ichimoto K., Katsukawa Y., Tsuneta S., Suematsu Y., Shimizu T., Nagata S.,
2008, {\it{\apj}}, \textbf{672}, 1237.

\bibitem[\protect\citeauthoryear{{Manso Sainz, Mart\`inez Gonz\'alez \& Asensio Ramos}}{2011}]{manso}
Manso Sainz R., Mart\`inez Gonz\'alez M. J., Asensio Ramos, A.,
2011, {\it{\aap}}, \textbf{531}, L9.

\bibitem[\protect\citeauthoryear{{Nordlund, Stein \& Asplund}}{2009}]{Nordlund2009}
Nordlund \AA{}., Stein R. F., Asplund M,. 
2009, {\it{Living Rev. Solar Phys}}, \textbf{6}, 2 [on line article 2011].

\bibitem[\protect\citeauthoryear{{November \& Simon}}{1988}]{November88}
November, L.~J. and Simon, G.~W.,
1988, {\it{\apj}}, \textbf{333}, 442.

\bibitem[\protect\citeauthoryear{{Rast}}{2003}]{rast}
Rast M.P.
2003, {\it{\apjl}}, \textbf{597}, 1200.

\bibitem[\protect\citeauthoryear{{Rees \& Semel}}{1979}]{reessemel}
Rees D. E. \& Semel M. D.,
1979, {\it{\aap}}, \textbf{74}, 1.

\bibitem[\protect\citeauthoryear{{S\`anchez Almeida}}{2003}]{almeida}
S\`anchez Almeida J.,
2003, {\it{\aap}}, \textbf{411}, 615.

\bibitem[\protect\citeauthoryear{{Solanki et al.}}{1996}]{solankietal}
Solanki S. K., Zufferey D., Lin H., Rüedi I., \& Kuhn J. R. 
1996, {\it{\aap}}, \textbf{310}, L33.

\bibitem[\protect\citeauthoryear{{Thomas}}{1990}]{thomas}
Thomas J. H.,
1990, {\it{Geophysical Monograph Series}}, \textbf{58}, 133.

\bibitem[\protect\citeauthoryear{{Trujillo Bueno}}{2003}]{truillo03} 
Trujillo Bueno, J.,
2003, in {\it Modelling of Stellar Atmospheres} Ed. by N. Piskunov, W.W. Weiss, and D. F. Gray, Published on behalf of the IAU by the Astronomical Society of the Pacific, 243.

\bibitem[\protect\citeauthoryear{{Tsuneta et al.}}{2008}]{tsuneta}
Tsuneta S., Ichimoto K., Katsukawa Y., Nagata S., Otsubo M., Shimizu T., Suematsu Y., Nakagiri M., Noguchi M., Tarbell T., and 15 coauthors,
2008, {\it{\solphys}}, \textbf{249}, 167.

%\bibitem[[\protect\citeautorsyear{{Vettolani et~al.}}{1985)}]{Vettolani85}
%Vettolani G., de Souza R. E., Marano B., Chincarini G. 
%1985, {\it{\aap}}, \textbf{144}, 506.

%\bibitem[\protect\citeauthoryear{{Ryden \& Melott}}{1996}]{Ryden96}
%Ryden BS & Melott AL 
%1996, {\it{\apj}}, \textbf{470}, 160.

\bibitem[\protect\citeauthoryear{{Uritsky \& Davila}}{2011}]{Uritsky}	
Uritsky V. M., Davila J. M.
2011, {\it arXiv:1111.5053} 

\bibitem[\protect\citeauthoryear{{Yakhot}}{1989}]{yakhot}
Victor Y,
1989, {\it Phys. Rev. L.}, \textbf{63}, 1965.

\bibitem[\protect\citeauthoryear{{Yelles Chaouche et al.}}{2011}]{Chaouche}
Yelles Chaouche L., Moreno-Insertis F., Mart\'inez Pillet V., Wiegelmann T., Bonet J. A., Knolker M., Bellot Rubio L. R., del Toro Iniesta J. C., Barthol P., Gandorfer A., Schmidt W., Solanki S. K.,
2011, {\it{\apjl}}, \textbf{727}, L30.


%-------------------------------------------------------------
 
%  
\end{thebibliography}
\end{document}